\documentclass[runningheads]{llncs}
\usepackage{graphicx}
\usepackage{multirow}
\PassOptionsToPackage{hyphens}{url}\usepackage{hyperref}
\begin{document}
\title{nnU-Net for Brain Tumor Segmentation}
\author{Fabian Isensee\inst{1} \and
Paul F. J\"ager\inst{1} \and
Peter M. Full\inst{1} \and
Philipp Vollmuth\inst{2}\thanks{né Kickingereder} \and
Klaus H.Maier-Hein\inst{1}}

\authorrunning{Isensee et al.}
\institute{Division of Medical Image Computing, German Cancer Research Center (DKFZ), Heidelberg, Germany \and
Section for Computational Neuroimaging, Department of Neuroradiology, Heidelberg University Hospital, Heidelberg, Germany\\
\email{f.isensee@dkfz-heidelberg.de}}
\maketitle              % typeset the header of the contribution
\begin{abstract}
We apply nnU-Net to the segmentation task of the BraTS 2020 challenge. The unmodified nnU-Net baseline configuration already achieves a respectable result. By incorporating BraTS-specific modifications regarding postprocessing, region-based training, a more aggressive data augmentation as well as several minor modifications to the nnU-Net pipeline we are able to improve its segmentation performance substantially. We furthermore re-implement the BraTS ranking scheme to determine which of our nnU-Net variants best fits the requirements imposed by it. Our method took the first place in the BraTS 2020 competition with Dice scores of 88.95, 85.06 and 82.03 and HD95 values of 8.498,17.337 and 17.805 for whole tumor, tumor core and enhancing tumor, respectively.

\end{abstract}
\section{Introduction}
Brain tumor segmentation is considered one of the most difficult segmentation problems in the medical domain. At the same time, widespread availability of accurate tumor delineations could significantly improve the quality of care by supporting diagnosis, therapy planning and therapy response monitoring \cite{kickingereder2019automated}. Furthermore, segmentation of tumors and associated subregions allows for identification of novel imaging biomarkers, which in turn enable more precise and reliable disease stratification \cite{kickingereder2018radiomic} and treatment response prediction \cite{kickingereder2016large}.

The Brain Tumor Segmentation Challenge (BraTS) \cite{menze2014multimodal,bakas2018identifying} provides the largest fully annotated and publicly available database for model development and is the go-to competition for objective comparison of segmentation methods. The BraTS 2020 dataset \cite{bakas2017advancing,bakas2018identifying,bakas2017segmentation,bakas2017segmentation2} comprises 369 training and 125 validation cases. Reference annotations for the validation set are not made available to the participants. Instead, participants can use the online evaluation platform\footnote{\url{https://ipp.cbica.upenn.edu/}} to evaluate their models and compare their results with other teams on the online leaderboard\footnote{\url{https://www.cbica.upenn.edu/BraTS20/lboardValidation.html}}. Besides the segmentation task, the BraTS 2020 competition features a survival prediction task as well as an uncertainty modelling task. In this work, we only participate in the segmentation task.

Recent successful entries in the BraTS challenge are exclusively based on deep neural networks, more specifically on encoder-decoder architectures with skip connections, a pattern which was first introduced by the U-Net \cite{ronneberger2015u}. Numerous architectural improvements upon the U-Net have been introduced, many of which are also used in the context of brain tumor segmentation, for example the addition of residual connections \cite{he2016deep,myronenko20183d,jiang2019two,isensee2017brain,wang2017automatic}, densely connected layers \cite{huang2017densely,zhao2019bag,mckinley2018ensembles,mckinley2019triplanar,mckinley2018ensembles} and attention mechanisms \cite{vaswani2017attention,mckinley2019triplanar}. In the context of network architectures, the winning contributions of 2018 \cite{myronenko20183d} and 2019 \cite{jiang2019two} should be highlighted as they extend the encoder-decoder pattern with a second decoder trained on an auxiliary task for regularization purposes. Training schemes are usually adapted to cope with the particular challenges imposed by the task of brain tumor segmentation. The stark class imbalance, for instance, necessitates appropriate loss functions for optimization: Dice loss \cite{milletari2016v,drozdzal2016importance} and focal loss \cite{lin2017focal} are popular choices \cite{isensee2018no,isensee2017brain,zhao2019bag,mckinley2019triplanar,myronenko20183d,jiang2019two,kamnitsas2017ensembles}. Since BraTS evaluates segmentations using the partially overlapping whole tumor, tumor core and enhancing tumor regions \cite{menze2014multimodal}, optimizing these regions instead of the three provided class labels (edema, necrosis and enhancing tumor) can be beneficial for performance \cite{wang2017automatic,isensee2018no,jiang2019two,zhao2019bag}. Quantifying the uncertainty in the data has also been shown to improve results \cite{mckinley2019triplanar}.

The methods presented above are highly specialized for brain tumor segmentation and their development required expertise as well as extensive experimentation. We recently proposed nnU-Net \cite{isensee2019automateddesign}, a general purpose segmentation method that automatically configures segmentation pipelines for arbitrary biomedical datasets. nnU-Net set new state of the art results on the majority of the 23 datasets it was tested on, underlining the effectiveness of this approach. In the following, we will investigate nnU-Net's suitability for brain tumor segmentation. We use nnU-Net both as a baseline algorithm and as a framework for model development.

\section{Method}

\subsection{Rankings should be used for model selection}
\label{use_the_ranking}
Optimizing a model for a competition is often mistakenly treated equivalently to optimizing the model towards the segmentation metrics used in that competition. Segmentation metrics, however, only tell half the story: they describe the model on a per-image level whereas the actual ranking is based on a consolidation of these the metrics across all test cases. Ranking schemes can be differentiated in 'aggregate then rank' and 'rank then average' approaches. In the former, some aggregated metric (for example the average) is computed and then used to rank the participants. In the latter, the participants are ranked on each individual training case and then their ranks are accumulated across all cases. Different algorithm characteristics may be desired depending on the ranking scheme that is used to evaluate them. For example, in an 'aggregate then rank' scheme, median aggregation (as opposed to the mean) would be more forgiving to algorithms that produce badly predicted outliers. We refer to \cite{maier2018rankings} for a comprehensive analysis on the impact of ranking on challenge results.

BraTS uses a 'rank then aggregate' approach, most likely because it is well suited to combine different types of segmentation metrics (such as HD95 and Dice). Effectively, each submission obtains six ranks per test case (one for each of the 3 evaluated regions times the 2 segmentation metrics) and the ranks are then averaged across all cases and metrics (see \footnote{\url{https://zenodo.org/record/3718904}}). The final rank is normalized by the number of participating algorithms to form the ranking score, which ranges from 0 (best) to 1 (worst).

The BraTS evaluation of cases with empty reference segmentations for enhancing tumor is tailored for the ranking scheme used by the competition: if a participant predicts false positive voxels in these cases, BraTS assigns the worst possible value for both metrics (Dice 0, HD95 373.13) whereas a correctly returned empty segmentation yields perfect values (Dice 1, HD95 0). Thus, the enhancing tumor label essentially becomes binary for all participants: they either achieve the (shared) first or (shared) last rank. 

As a consequence, optimizing models to maximize the mean-aggregated Dice scores and HD95 values returned by the BraTS evaluation platform may not be an ideal surrogate for optimal performance on the BraTS competition. We therefore reimplemented the ranking used by the BraTS competition and used it to rank our models against each other in order to select the best performing variant(s).

\subsection{nnU-Net baseline}

We base our method on nnU-Net, our recently proposed fully automatic framework for the configuration of segmentation methods. We refer to \cite{isensee2019automateddesign} for a detailed description of nnU-Net.

\begin{figure}
\includegraphics[width=\textwidth]{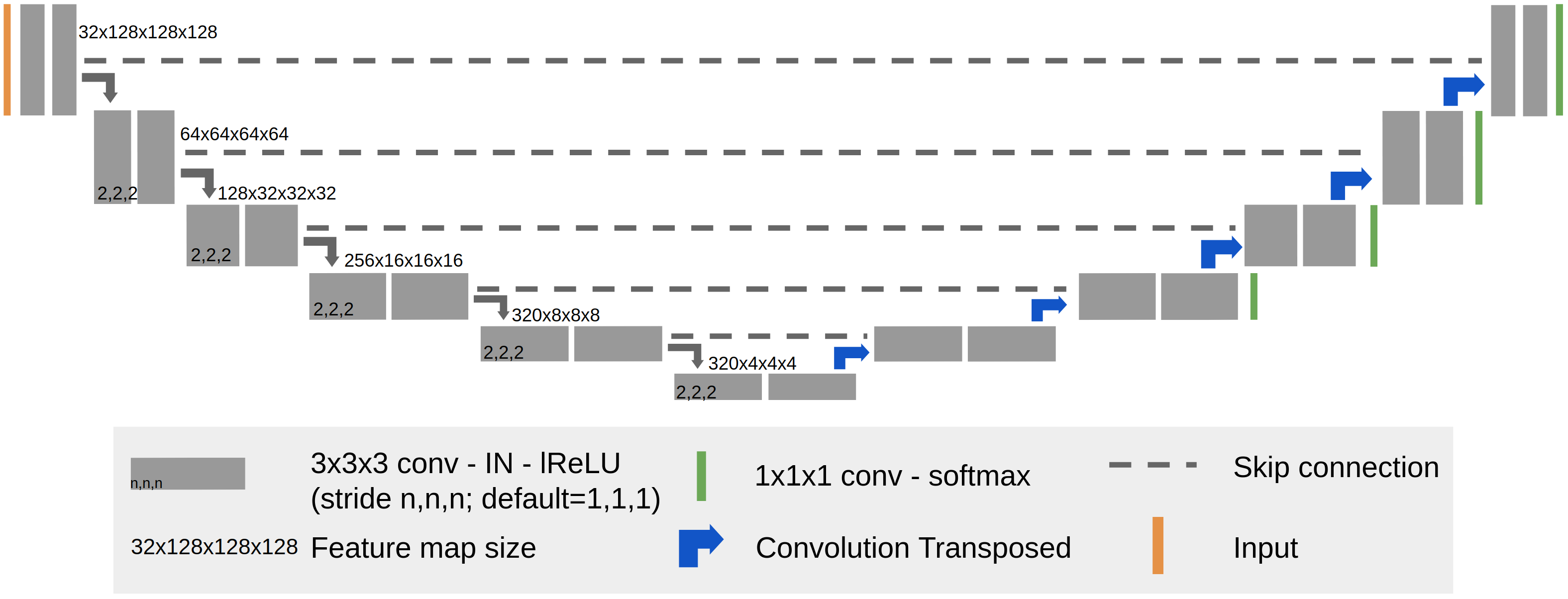}
\caption{Network Architecture as generated by nnU-Net. nnU-Net uses only plain U-Net-like architectures. For BraTS 2020, an input patch size of $128\times128\times128$ was selected. Downsampling is done with strided convolutions, upsampling is implemented as convolution transposed. Feature map sizes are displayed in the encoder part of the architecture. The feature maps in the decoder mirror the encoder. Auxiliary segmentation outputs used for deep supervision branch off at all but the two lowest resolutions in the decoder.} \label{nnunet_arch}
\end{figure}

\label{nnunet_baseline}
First, we apply nnU-Net without any modifications to provide a baseline for later modifications. The design choices made by nnU-Net are described in the following.

nnU-Net normalizes the brain voxels of each image by subtracting their mean and dividing by their standard deviation. The non-brain voxels remain at 0. The network architecture generated by nnU-Net is displayed in Figure \ref{nnunet_arch}. It follows a 3D U-Net\cite{cciccek20163d,ronneberger2015u} like pattern and consists of an encoder and a decoder which are interconnected by skip connections. nnU-net does not use any of the recently proposed architectural variations and only relies on plain convolutions for feature extraction. Downsampling is performed with strided convolutions and upsampling is performed with convolution transposed. Auxiliary segmentation outputs, which are used for deep supervision, branch off at all but the two lowest resolutions in the decoder. The input patch size is selected to be $128\times128\times128$ with a batch size of 2. A total of five downsampling operations are performed, resulting in a feature map size of $4\times4\times4$ in the bottleneck. Initial number of convolutional kernels is set to 32, which is doubled with each downsampling up to a maximum of 320. The number of kernels in the decoder mirrors the encoder. Leaky ReLUs \cite{maas2013rectifier} are used as nonlinearities. Instance normalization \cite{ulyanov2016instance} is used for feature map normalization.

Training objective is the sum of Dice \cite{drozdzal2016importance,milletari2016v} and cross-entropy loss. The loss operates on the three class labels edema, necrosis and enhancing tumor. nnU-Net uses stochastic gradient descent with an initial learning rate of 0.01 and a Nesterov momentum of 0.99. Training runs for a total of 1000 epochs, where one epoch is defined as 250 iterations. The learning rate is decayed with a polynomial schedule as described in \cite{chen2017deeplab}. Training patches are cropped from randomly selected training cases. Data augmentation is applied on the fly during training (see also Supplementary Information D in \cite{isensee2019automateddesign}).

\subsection{BraTS-specific optimizations}
\label{optimize_nnunet_for_brats}
Besides its role as a high quality standardized baseline and out-of-the-box segmentation tool, we also advertise nnU-Net as a framework for method development. To underline this aspect of nnU-Net's broad functionality, we select promising BraTS-specific modifications and integrate them into nnU-Net's configuration.

\subsubsection{Region-based training}
\label{region_training}
The provided labels for training are 'edema', 'non-enhancing tumor and necrosis' and 'enhancing tumor'. The evaluation of the segmentations is, however, performed on three partially overlapping \textit{regions}: whole tumor (consisting of all 3 classes), tumor core (non-enh. \& necrosis + enh. tumor) and enhancing tumor. It has been shown previously \cite{wang2017automatic,isensee2018no,jiang2019two,zhao2019bag,mckinley2019triplanar,myronenko20183d} that directly optimizing these regions instead of the individual classes can improve performance on the BraTS challenge. To this end, we replace the softmax nonlinearity in our network architecture with a sigmoid and change the optimization target to the three tumor subregions. We also replace the crossentropy loss term with a binary cross-entropy that optimizes each of the regions independently.

\subsubsection{Postprocessing}
When the reference segmentation for enhancing tumor is empty, BraTS evaluation awards zero false positive predictions with a Dice score of 1 (the Dice would otherwise be undefined due to division by 0), thus placing the corresponding algorithm in the (shared) first rank for this test case, region and metric. This can be exploited with the goal of improving the average rank of the submitted method and thus its overall ranking in the challenge. By removing enhancing tumor entirely if the predicted volume is less than some threshold, one can accumulate more perfect rankings at the expense of some additional cases with a Dice score of 0 (and corresponding worst rank). Even though this strategy has the side effect of removing some true positive predictions, the net gain can out-weigh the losses. Removed enhancing tumor is replaced with necrosis to ensure that these voxels are still considered part of the tumor core. We optimize the threshold for postprocessing on our training set cross-validation twice, once via maximizing the mean Dice score and once via minimizing the ranking score in our internal BraTS-like ranking. Whenever we present postprocessed results, we select the best value even if that value was achieved by the opposing selection strategy.

\subsection{Further nnU-Net Modifications}
\label{further_nnunet_modifications}
\subsubsection{Increased batch size}
Over the past years, the BraTS dataset has continued to grow in size. The low batch size used by nnU-Net results in noisier gradients, which potentially reduces overfitting but also constrains how accurately the model can fit the training data. With a larger dataset, it may be beneficial to increase the batch size (bias variance trade-off). To this end, we modify nnU-Net's configuration to increase the batch size from 2 to 5 in an attempt to improve the model accuracy.

\subsubsection{More data augmentation}
\label{moreDA}
Data augmentation can effectively be used to artificially enlarge the training set. While nnU-Net already uses a broad range of aggressive data augmentation techniques, we make use of even more aggressive augmentations in an attempt to increase the robustness of our models. All augmentations are applied on the fly during training using the \textit{batchgenerators} framework\footnote{\url{https://github.com/MIC-DKFZ/batchgenerators}}. Relative to the nnU-Net baseline, we make the following changes:
\begin{itemize}
    \item increase the probability of applying rotation and scaling from 0.2 to 0.3.
    \item increase the scale range from (0.85, 1.25) to (0.65, 1.6)
    \item select a scaling factor for each axis individually
    \item use elastic deformation with a probability of 0.3
    \item use additive brightness augmentation with a probability of 0.3
    \item increase the aggressiveness of the Gamma augmentation
\end{itemize}

\subsubsection{Batch normalization}
In our participation in the M\&Ms challenge \footnote{\url{https://www.ub.edu/mnms/}} we noticed that more aggressive data augmentation could be used to effectively close the domain gap to other scanners, but only when used in conjunction with batch normalization (instead of instance normalization) (results available here \footnote{\url{https://www.ub.edu/mnms/results.html}}, our paper is not yet available). In BraTS, Dice scores for the test cases are often lower than the reported values on the training and validation dataset, which makes us believe that there may be a domain gap between the test set and the training and validation sets. This suggests that pursuing a this strategy for BraTS as well may be beneficial. 

\subsubsection{Batch dice}
\label{batch_dice}
The standard implementation of the Dice loss computes the loss for every sample in the minibatch independently and then averages the loss over the batch (we call this the \textit{sample Dice}). Small errors in samples with few annotated voxels can cause large gradients and dominate the parameter updates during training. If these errors are caused by imperfect predictions of the model, these large gradients are desired to push the model towards better predictions. However, if the models predictions are accurate and the reference segmentation is imperfect, these large gradients will be counterproductive during training. We therefore implement a different Dice loss computation: instead of treating the samples in a minibatch independently, we compute the dice loss over all samples in the batch (pretending they are just a single large sample, we call this the \textit{batch Dice}). This effectively regularizes the Dice loss because samples with previously only few annotated voxels are now overshadowed by other samples in the same batch. 

\subsubsection{Abbreviations}
\label{abbreviations}
We will represent our models using the following abbreviations:

\begin{itemize}
    \item \textbf{BL/BL*}: baseline nnU-Net without modifications. * indicates a batch size of 5 instead of 2.
    \item \textbf{R}: Region-based training (see Section \ref{region_training})
    \item \textbf{DA/DA*}: more aggressive data augmentation as described in Section \ref{moreDA}. * indicates that the brightness augmentation is only applied with a probability of 0.5 for each input modality.
    \item \textbf{BD}: model trained with batch Dice (as opp posed to sample Dice, see Section \ref{batch_dice})
\end{itemize}

We denote our models by the modifications that were applied to it, for example BL*+R+BD. Note that all models shown will have postprocessing applied to them.

\section{Results}
\subsection{Aggregated Scores}
\begin{table}[]
\centering
\caption{Training set results (n=369). All experiments were run as 5-fold cross-validation. See Section \ref{abbreviations} for decoding the abbreviations}
\begin{tabular}{l|ccc|c}
\multirow{2}{*}{Model} & \multicolumn{4}{l}{Training Set (Dice, 5-fold CV)} \\
 & Whole & Core & Enh. & Mean \\ \hline
BL & 91.6 & 87.23 & 80.83 & 86.55 \\
BL* & 91.85 & 86.24 & 80.18 & 86.09 \\
BL*+R & 91.75 & 87.24 & 82.21 & 86.73 \\
BL*+R+DA & 91.87 & 87.97 & 81.37 & 87.07 \\
BL*+R+DA+BN & 91.57 & 87.59 & 81.29 & 86.82 \\
BL*+R+DA+BD & 91.76 & 87.67 & 80.94 & 86.79 \\
BL*+R+DA+BN+BD & 91.7 & 87.21 & 81.7 & 86.87 \\
BL*+R+DA*+BN & 91.6 & 87.51 & 80.94 & 86.68 \\
BL*+R+DA*+BN+BD & 91.47 & 87.13 & 81.33 & 86.64
\end{tabular}
\label{trainsetresults}
\end{table}

We train each configuration as a five-fold cross-validation on the training cases (no external data is used). This not only provides us with a performance estimate on the 369 training cases but also enables us to select the threshold for postprocessing on the training set rather than the validation set. The results for the different configurations are presented in Table \ref{trainsetresults}.

\begin{table}[]
\centering
\caption{Validation set results. Predictions were made using the 5 models from the training cross-validation as an ensemble. Metrics computed by the validation platform. See Section \ref{abbreviations} for decoding the abbreviations}
\begin{tabular}{l|ccc|c||ccc|c}
\multirow{2}{*}{Model} & \multicolumn{4}{c}{Dice} & \multicolumn{4}{c}{HD95} \\
 & Whole & Core & Enh. & Mean & Whole & Core & Enh. & Mean \\ \hline
BL & 90.6 & 84.26 & 77.67 & 84.18 & 4.89 & 5.91 & 35.10 & 15.30 \\
BL* & 90.93 & 83.7 & 76.64 & 83.76 & 4.23 & 6.01 & 41.06 & 17.10 \\
BL*+R & 90.96 & 83.76 & 77.65 & 84.13 & 4.41 & 8.80 & 29.82 & 14.34 \\
BL*+R+DA & 90.9 & 84.61 & 78.67 & 84.73 & 4.70 & 5.62 & 29.50 & 13.28 \\
BL*+R+DA+BN & 91.24 & 85.04 & 79.32 & 85.2 & 3.97 & 5.17 & 29.25 & 12.80 \\
BL*+R+DA+BD & 90.97 & 83.91 & 77.48 & 84.12 & 4.11 & 8.60 & 38.06 & 16.93 \\
BL*+R+DA+BN+BD & 91.15 & 84.19 & 79.99 & 85.11 & 3.72 & 7.97 & 26.28 & 12.66 \\
BL*+R+DA*+BN & 91.18 & 85.71 & 79.85 & 85.58 & 3.73 & 5.64 & 26.41 & 11.93 \\
BL*+R+DA*+BN+BD & 91.19 & 85.24 & 79.45 & 85.29 & 3.79 & 7.77 & 29.23 & 13.60
\end{tabular}
\label{valsetresults}
\end{table}

We use the five models obtained from the cross-validation on the training cases as an ensemble to predict the validation set. The aggregated Dice scores and HD95 values as computed by the online evaluation platform are reported in Table \ref{valsetresults}. Again we provide averages of the HD95 values and Dice scores across the three evaluated regions for clarity. 

As discussed in  Section \ref{use_the_ranking}, these aggregated metric should not be used to run model selection for the BraTS because they may not directly optimize the ranking that will be used to evaluate the submissions. This becomes evident when looking at the HD95 values of the enhancing tumor HD95. The HD95 averages are artificially inflated because false positive predictions in cases where no reference segmentation for enhancing tumor is present are hard coded to receive a HD95 value of 373.13. Given that the vast majority of HD95 for this class is much smaller (even the 90th percentile is just 11.32 for our final validation set submission), these outlier values dominate the mean aggregation and result in mostly uninformative mean HD95: Switching the metric a single case from 373.13 to 0 (due to the removal of false positive predictions) can have a huge effect on the mean. The same is true for the Dice score, but to a lesser extend: a HD95 of 373.13 is basically impossible to occur even if the predicted segmentation is of poor quality whereas a low Dice score is likely to happen.

\subsection{Internal BraTS-like ranking}
\begin{table}[]
\centering
\caption{Rankings of our proposed nnU-Net variants on the training and validation set. For each set, we rank twice: once based on the mean Dice and once using our reimplementation of the BraTS ranking.}
\begin{tabular}{l|cc|cc||cc|cc}
\multirow{2}{*}{Model} & \multicolumn{4}{c||}{Training set} & \multicolumn{4}{c}{Validation set} \\
 & \multicolumn{2}{c|}{\begin{tabular}[c]{@{}c@{}}rank based\\ on mean Dice\end{tabular}} & \multicolumn{2}{c||}{BraTS ranking} & \multicolumn{2}{c|}{\begin{tabular}[c]{@{}c@{}}rank based\\ on mean Dice\end{tabular}} & \multicolumn{2}{c}{BraTS ranking} \\
 & value & rank & value & rank & value & rank & value & rank \\ \hline
BL & 86.55 & 8 & 0.3763 & 8 & 84.18 & 6 & 0.4079 & 8 \\
BL* & 86.09 & 9 & 0.3767 & 9 & 83.76 & 9 & 0.4236 & 9 \\
BL*+R & 86.73 & 5 & 0.3393 & 5 & 84.13 & 7 & 0.4005 & 7 \\
BL*+R+DA & 87.07 & \textbf{1} & 0.3243 & 3 & 84.73 & 5 & 0.3647 & 5 \\
BL*+R+DA+BN & 86.82 & 3 & 0.3377 & 4 & 85.20 & 3 & 0.3577 & 4 \\
BL*+R+DA+BD & 86.79 & 4 & 0.3231 & 2 & 84.12 & 8 & 0.3726 & 6 \\
BL*+R+DA+BN+BD & 86.87 & 2 & 0.3226 & \textbf{1} & 85.11 & 4 & 0.3487 & 3* \\
BL*+R+DA*+BN & 86.68 & 6 & 0.3521 & 6 & 85.58 & \textbf{1} & 0.3125 & \textbf{1*} \\
BL*+R+DA*+BN+BD & 86.64 & 7 & 0.3595 & 7 & 85.29 & 2 & 0.3437 & 2*
\end{tabular}
\label{ranking_results}
\end{table}

To overcome the shortcomings of metric aggregation-based model selection we create an internal leaderboard where we rank our nnU-Net variants presented above (as well as others that we omitted for brevity) against each other using a reimplementation of the BraTS ranking. Results of both ranking schemes are summarized in Table \ref{ranking_results}. While the ranks between the mean Dice and the BraTS ranking scheme are different, both still seem to follow similar trends. On the training set, for example, the BL and BL* models constitute the worst contenders while BL*+R+DA+BN+BD performs quite well. Within the validation set the correlation between the ranking schemes es even more pronounced. When comparing the model performance between the training and validation sets (any of the ranking schemes) the rankings differ drastically: the DA* models perform exceptionally well on the validation set (rank 1 and 2) whereas they underperform on the training set (rank 6 and 7). This creates a difficult situation. How do we choose the models used for the test set submission? Do we trust more in the 125 validation cases because we do not have access to the reference labels and thus can not overfit as easily or do we favor the training set because is has almost 3 times as many cases and should therefore provide more reliable performance estimates?

We opted for trusting the validation set over the training set and thus selected the three top performing models to build our final ensemble: $\mathrm{BL^*}+\mathrm{R}+\mathrm{DA^*}+\mathrm{BN}+\mathrm{BD}$, BL*+R+DA*+BN and BL*+R+DA+BN+BD. Note that additionally to the 5 models from the cross-validation we had 10 more models of the BL*+R+DA*+BN configuration (each trained on a random 80:20 split of the training cases). Thus, our final ensemble consisted of 5 + 5 + 15 = 25 models. Note that ensembling was implemented by first predicting the test cases individually with each configuration, followed by averaging the sigmoid outputs to obtain the final prediction. Therefore, the 15 models of the BL*+R+DA*+BN configuration had the same influence on the final prediction as the other configurations.

Our final ensemble achieved mean Dice scores of 91.24, 85.06 and 79.89 and HD95 of 3.69, 7.82 and 23.50 for whole tumor, tumor core and enhancing tumor, respectively (on the validation set). With a mean Dice score of 85.4 the ensemble appears to perform worse in terms of the 'aggregate then rank' approach. However, since it sits comfortably on the first place of our internal 'rank then aggregate' ranking with a score of 0.2963 we are confident in its segmentation performance.

\subsection{Qualitative results}

\begin{figure}
\centering
\includegraphics[width=0.81\textwidth]{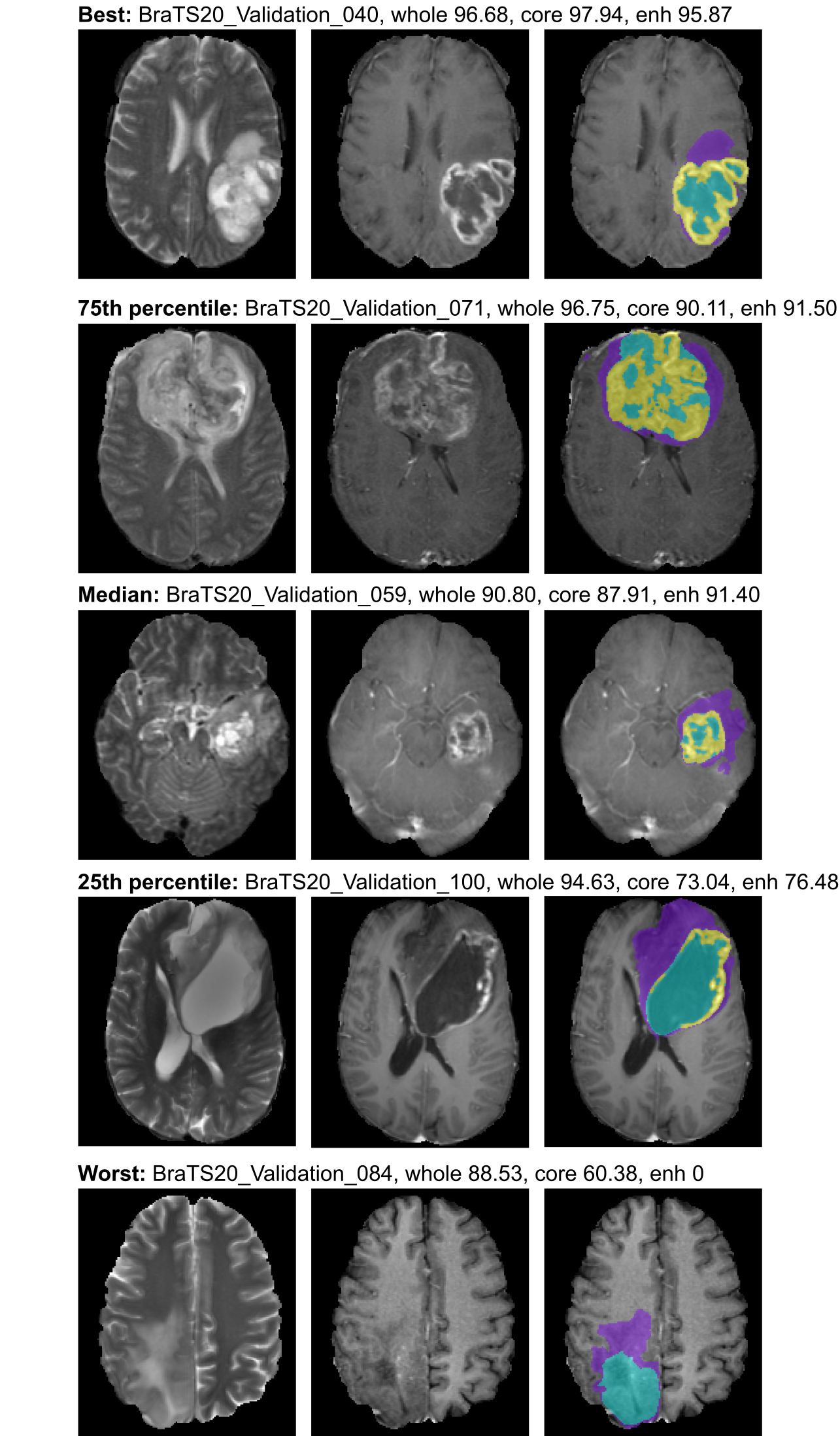}
\caption{Qualitative validation set results. Cases were selected as best, worst, median and 75th and 25th percentile. Within each row, the raw T2 image is shown to the left, the T1c image in the middle and on overlay with the generated segmentation on the T1c image is shown on the right. Edema is shown in violet, enhancing tumor in yellow and necrosis/non-enhancing tumor on turquoise.} \label{valresults}
\end{figure}

Figure \ref{valresults} provides a qualitative overview of the segmentation performance. It shows results generated by our ensemble on the validation set. To rule out cherry picking we standardized the selection of presented validation cases: They were selected as best, worst, median and 75th and 25th percentiles based on their Dice scores (averaged over the three validation regions). As can be seen in the figure, the segmentation quality is high overall. The low tumor core score for the \textit{worst} example hints at one of the potential issues with the definition of this class in the reference segmentations (see Discussion). The enhancing tumor score of 0 in the absence of predicted enhancing tumor voxels indicates either that our model missed a small enhancing tumor lesion or that it was removed as a result of our postprocessing. An inspection of the non-postprocessed segmentation mask reveals that the enhancing tumor lesion was indeed segmented by the model and must have been removed during postprocessing. 

\subsection{Test set results}
\begin{table}[]
\centering
\caption{Quantitative test set results. Values were provided by the challenge organizers.}
\begin{tabular}{l|ccc|ccc}
\multirow{2}{*}{} & \multicolumn{3}{c}{Dice} & \multicolumn{3}{c}{HD95} \\
 & Enh & Whole & Core & Enh & Whole & Core \\ \hline
Mean & 82.03 & 88.95 & 85.06 & 17.805 & 8.498 & 17.337 \\
StdDev & 19.71 & 13.23 & 24.02 & 74.886 & 40.650 & 69.513 \\
Median & 86.27 & 92.73 & 92.98 & 1.414 & 2.639 & 2.000 \\
25th percentile & 79.30 & 87.76 & 88.33 & 1.000 & 1.414 & 1.104 \\
75th percentile & 92.25 & 95.17 & 96.19 & 2.236 & 4.663 & 3.606
\end{tabular}
\label{testsetresults}
\end{table}

Table \ref{testsetresults} provides quantitative test set results. Our submission took the first place in the BraTS 2020 competition (see \url{https://www.med.upenn.edu/cbica/brats2020/rankings.html}).

\section{Discussion}

This manuscript describes our participation in the segmentation task of the BraTS 2020 challenge. We use nnU-Net \cite{isensee2019automateddesign} not only as our baseline algorithm but also as a framework for method development. nnU-Net again proved its generalizability by providing high segmentation accuracy in an out-of-the-box fashion. On the training set cross-validation, the performance of the nnU-Net baseline is very close to the configurations that were specifically modified for the BraTS challenge. On the validation set, however, the proposed BraTS-specific modifications provide substantially higher segmentation performance. Our final ensemble was selected based on the three best models on the validation set (as determined with our reimplementation of the BraTS ranking). It obtained validation Dice scores of 91.24, 85.06 and 79.89 as well as HD95 of of 3.69, 7.82 and 23.5 for whole tumor, tumor core and enhancing tumor, respectively. 

Our approach won the BraTS 2020 competition. On the test set, our model obtained Dice scores of 88.95, 85.06 and 82.03 and HD95 values of 8.498, 17.337 and 17.805 for whole tumor, tumor core and enhancing tumor, respectively. 

We should note that this manuscript only spans a small number of modifications and lacks sufficiently extensive experimental validation thereof. While most of our nnU-Net configurations achieve good performance, our results do not allow us to accurately pinpoint which aspects contributed the most. Some design choices may even have reduced our overall performance. Increasing the batch size from 2 to 5 gave worse results according to both evaluated ranking schemes and on both the training and validation set (see Table \ref{ranking_results}). This is a clear signal that the batch size should have been reverted to 2 for the remaining experiments. We also believe that a more thorough optimization of hyperparameters regarding both the training scheme as well as the data augmentation could result in further performance gains. 

Based on our observations in Figure \ref{valresults}, one major source of error appears to be the tumor core prediction. In particular the \textit{worst} presented example has a rather low Dice score for this region, even though the result seems visually plausible. We believe that this failure mode is not necessarily an issue inherent to our model but potentially originates from an inconsistency in the definition of the \textit{non-enhancing tumor and necrosis} label, particularly in LGG cases. While the necrosis part of this label is easy to recognize, the non-enhancing tumor region often has little evidence in the image and the associated annotations may be subjective. 

The enhancing tumor class is arguably the most difficult to segment in this dataset. What makes this class particularly challenging is the way its evaluation is handled when the reference segmentation of an image does not contain this class. The BraTS evaluation scheme favors the removal of small enhancing tumor lesions and thus encourages such postprocessing. In a clinical scenario where the accurate detection of small enhancing tumors could be critical, this property is not necessarily desired and we recommend to omit the postprocessing presented in this manuscript.

Just like many other BraTS participants, we have used the aggregated Dice scores and HD95 values for model selection in previous BraTS challenges \cite{isensee2017brain,isensee2018no}. As discussed in this manuscript, this strategy may not be ideal because the ranking that is used to determine the winning contributions works on a different principle. Indeed, when comparing the rankings of our models relative to each other on the training and validation set, we observed disparities between 'aggregate then rank' and 'rank then aggregate'. However, the overall trends were similar between the ranking methods indicating that a model selection based on metric aggregation is still going into the right direction. 

While experimenting with the 'rank then aggregate' ranking on our model configurations we noted some instability in the ranks obtained by out models: when models are added or removed from the pool of methods, the ranking of other methods relative to each other could change. This seems to be an inherent property of this ranking  \cite{maier2018rankings}. Even though this might appear alarming at first glance, we attribute most of the instability to the similarity of the methods within the pool which has certainly overemphasized this potential issue. We expect the ranking to be more stable the more diverse the evaluated methods are (especially on the test set where all methods are compared against each other).

Given that aggregated metric are at least in part disconnected from the actual evaluation of the challenge, it is frustrating that the online leaderboards for training and validation sets only display mean values for the competing methods and therefore do not allow for a perfectly accurate comparison with other participants. Especially in cases where the enhancing tumor class is absent from the reference segmentation, the dichotomy of the returned metric values (0 or 1 for the Dice score and 0 or 373.13 for the HD95) can overshadow the more minute (but still very meaningful) differences between the teams on the validation cases that do actually contain this label. We would love to see the 'rank then aggregate' ranking scheme implemented in the online leaderboards as well so that they become more informative about a models performance on the BraTS challenge. 

The source code for our submission is publicly available as part of the nnU-Net repository (\url{https://github.com/ MIC-DKFZ/nnUNet}). A docker image for reproducing our test set predictions is available on Docker Hub\footnote{\url{https://hub.docker.com/repository/docker/fabianisensee/isen2020}}.

\bibliographystyle{IEEEtran}
\bibliography{bibliography}
	
\end{document}